\documentstyle[12pt]{article}
\advance\textheight by \topskip
\makeatletter

\@addtoreset{equation}{section}
\makeatother

\newcommand{\BE}{\begin{equation}}
\newcommand{\EE}{\end{equation}}
\newcommand{\BA}{\begin{eqnarray}}
\newcommand{\EA}{\end{eqnarray}}

\def\no{\nonumber}
\def\bi{\bibitem}

\def\ap{\alpha '}
\def\B{\beta}
\def\BH{\beta_{\scriptscriptstyle H}}
\def\T2{|T|^2}
\def\TM{{\bf T}}
\def\p{\partial}
\def\D#1{D#1-$\overline{\textrm{D#1}}$}
\def\v#1{{\cal V}_{#1}}
\def\vp{{\cal V}_{p}}

\def\DD{D-brane$-$anti-D-brane}
\def\F{{\cal F}}

\input epsf

\begin{document}

\rightline{KEK-TH-1068}
\rightline{hep-th/0601133}

\vspace{.8cm}
\begin{center}
{\large\bf Non-BPS D9-branes in the Early Universe
}

{\bf Kenji Hotta,}
\footnote{E-mail address: khotta@post.kek.jp}

{\it High Energy Accelerator Research Organization (KEK),

Tsukuba, Ibaraki, 305-0801, Japan}

\vskip 1.5cm

\end{center}
\vskip .6 cm
\centerline{\bf ABSTRACT}
\vspace{-0.7cm}
\begin{quotation}

We have investigated the finite temperature systems of non-BPS D-branes and {\DD} pairs in the previous papers. It has been shown that non-BPS D9-branes and \D{9} pairs become stable near the Hagedorn temperature on the basis of boundary string field theory. This implies that there is a possibility that these spacetime-filling branes exist in the early universe. We study the time evolution of the universe in the presence of non-BPS D9-branes on the basis of boundary string field theory in this paper. We try to construct the following scenario for the early universe: The universe expands at high temperature and the open string gas on the non-BPS D9-branes dominates the total energy of the system at first. The temperature decreases as the universe expands. Then the non-BPS D9-branes become unstable at low temperature and decay through tachyon condensation. We obtain some classical solutions for Einstein gravity and dilaton gravity in the very simple cases.

\end{quotation}

\normalsize
\newpage

\section{Introduction}
\label{sec:Intro}

Non-BPS D-branes and {\DD} pairs are unstable systems in superstring theory \cite{nonBPSD} (for a review see, e.g., Ref. \cite{Ohmori}). Type IIB string theory contains non-BPS D-branes of even dimension and {\DD} pairs of odd dimension, whereas type IIA string theory contains non-BPS D-branes of odd dimension and {\DD} pairs of even dimension. The spectrum of open strings on these unstable branes contains a tachyon field $T$. In such a brane configuration, we have $T=0$, and the potential of this tachyon field has a local maximum at $T=0$. If we assume that the tachyon potential has a non-trivial minimum, it is hypothesized that the tachyon falls into it. Sen conjectured that the potential height of the tachyon potential exactly cancels the tension of the original unstable D-branes \cite{Senconjecture}. This implies that these unstable brane systems disappear at the end of the tachyon condensation.

We have pointed out in the previous papers that there are the cases that these unstable branes become stable at sufficiently high temperature \cite{Hotta4} \cite{Hotta5} \cite{Hotta6}\footnote{For related discussions see Refs. \cite{LowTtach} and \cite{Huang}.}. In particular, the spacetime-filling branes such as non-BPS D9-branes and {\D{9}} pairs become stable near the Hagedorn temperature in all the cases we have studied. The aim of this paper is to apply these works to cosmology. We will investigate the time evolution of the universe in the presence of non-BPS D9-branes.

We can calculate the tachyon potential for these unstable branes by using boundary string field theory (BSFT) \cite{BSFT1} \cite{BSFT2}. The tree level tachyon potential of $N$ non-BPS D-branes is given by \cite{tachyon2} \cite{TakaTeraUe}\footnote{We adopt the natural unit $c = \hbar = 1$.}
\BE
  V(\TM) = \sqrt{2} \ \tau_p \vp \ \textrm{Tr} \ \exp (- \alpha {\TM}^2),
\EE
where $\TM$ is a real scalar field in the $(N, \overline{N})$ representation of the $U(N) \times U(N)$ gauge group \cite{IIBornotIIB}, $\tau_p$ is the tension of a single D$p$-brane, $\vp$ is the $p$-dimensional volume of the system that we are considering, and $\alpha$ is a constant which depends on the notation. $\tau_p$ is defined by
\BE
  \tau_p = \frac{1}{(2 \pi)^p {\ap}^{\frac{p+1}{2}} g_s},
\EE
where $g_s$ is the coupling of strings, which is represented as
\BA
  g_s = e^{\phi},
\EA
by using the dilaton $\phi$. $\ap$ is the slope parameter and we will set $\ap = 1$ when we perform numerical calculations. The potential has a local maximum at $\TM = 0$. It becomes minimum when $\textrm{Tr} \ \exp (- \alpha {\TM}^2) = 0$, and it satisfies Sen's conjecture. Let us suppose that $\TM$ has the following form;
\BE
  \TM =
     \left( \begin{array}{ccccc}
       T   &       &       &       &   0   \\
           & \cdot &       &       &       \\
           &       & \cdot &       &       \\
           &       &       & \cdot &       \\
       0   &       &       &       &   T   \\
     \end{array} \right).
\label{eq:TM}
\EE
Then the tachyon potential is given by
\BE
  V(T) = \sqrt{2} \ N \tau_p \vp \exp (- \alpha T^2).
\EE
The potential has a local maximum at $T=0$ and has the minimum at $|T| = \infty$ in this case. We will deal with only this type of matrix in this paper.

We summarize here the results of our previous works about the finite temperature systems of non-BPS D-branes and {\DD} pairs in a constant tachyon background \cite{Hotta4} \cite{Hotta5} \cite{Hotta6}. Since we can obtain the similar results both in the case of {\DD} pairs and non-BPS D-branes, we only review the non-BPS D-brane case here. We have computed the finite temperature effective potential by using the Matsubara method in the framework of BSFT in order to study the thermodynamical behavior of these systems. If we consider the one-loop amplitude based on BSFT, we are confronted with an ambiguity in the choice of the Weyl factors of the two boundaries of open string world sheet. We have used the boundary action which has been proposed by Andreev and Oft \cite{1loopAO}.

We have investigated non-BPS D$p$-branes in a non-compact flat 10-dimensional Minkowski background. The result of the non-BPS D9-brane case is in sharp contrast to that of the non-BPS D$p$-brane case with $p \leq 8$. In the case of $N$ non-BPS D9-branes, the $T^2$ term of the finite temperature effective potential near the Hagedorn temperature is approximated as
\BE
  \frac{\alpha}{8} \left[ -8 \sqrt{2} \ N \tau_9 \v{9}
   + \frac{4 \pi N^2 \v{9}}{{\BH}^{10}} \ 
    \ln \left( \frac{\pi {\BH}^{10} E}{N^2 \v{9}}
     \right) \right] T^2,
\EE
where $E$ is the energy of open strings. Here, we must impose the condition that the 't Hooft coupling is very small, namely,
\BE
  g_s N \ll 1,
\label{eq:gN}
\EE
for the one-loop approximation. Because the first term in the coefficient of $T^2$ is a constant as long as $\v{9}$ and $\tau_9$ are fixed and the second term is an increasing function of $E$, the sign of the $T^2$ term changes from negative to positive at large $E$. The coefficient vanishes when
\BA
  E_c &\simeq& \frac{N^2 \v{9}}{\pi {\BH}^{10}}
    \exp \left( \frac{2 \sqrt{2} {\BH}^{10} \tau_9}{\pi N} \right),
\label{eq:EnonBPS} \\
  {\cal T}_c
    &\simeq& {\BH}^{-1}
      \left[ 1 + \exp \left( - \ \frac{\sqrt{2} {\BH}^{10} \tau_9}
        {\pi N} \right) \right]^{-1}.
\EA
From this we can see that a phase transition occurs at the temperature ${\cal T}_c$, which is slightly below the Hagedorn temperature, and the non-BPS D9-branes are stable above this critical temperature. The total energy at the critical temperature is a decreasing function of $N$ as long as the 't Hooft coupling is very small in these cases. This implies that a large number $N$ of non-BPS D9-branes are created simultaneously.\footnote{We cannot determine the value of $N_{min}$ with a perturbative calculation. In order to determine it, we must perform a non-perturbative calculation based on, for example, the matrix model \cite{matrix} or the IIB matrix model \cite{IIBmatrix}. The K-matrix model may also be useful, as it explicitly contains the tachyon field \cite{Kmatrix}.} For the non-BPS D$p$-branes with $p \leq 8$, on the other hand, the coefficient remains negative near the Hagedorn temperature, so that such a phase transition does not occur. We thus concluded that not lower dimensional non-BPS D-branes but non-BPS D9-branes are created near the Hagedorn temperature.

We have also investigated non-BPS D$p$-branes in a toroidal background. We have supposed that $D$-dimensional space is compactified and that the rest of the $(9-D)$-dimensional space is left uncompactified ($M_{1,9-D} \times T_{D}$). We have also assume that the non-BPS D$p$-branes extend in the $d$-dimensional space in the non-compact direction and in the $(p-d)$-dimensional space in the toroidal direction. If $D+d=9$, that is, the non-BPS D$p$-branes are extended in all the non-compact directions, a phase transition occurs near the Hagedorn temperature and these branes become stable. On the other hand, if $D+d \leq 8$, that is, the non-BPS D$p$-branes are not extended in all the non-compact directions, such a phase transition does not occur. It is noteworthy that spacetime-filling branes, such as non-BPS D9-branes and {\D{9}} pairs, are created at sufficiently high energy not only in a non-compact background but also in a toroidal background, since these branes always satisfy $D+d=9$.

These results drive us to the question what happens if we apply the above arguments to cosmology. There is a possibility that non-BPS D-branes and {\DD} pairs exist in the early universe. We investigate the evolution of the universe in the presence of non-BPS D9-branes in this paper. We choose the non-BPS D9-branes because they become stable near the Hagedorn temperature in all the cases we have studied, and we can impose the homogeneous and isotropic condition if we consider such spacetime-filling branes\footnote{We investigate the non-BPS D9-brane case instead of the case of the lower-dimensional non-BPS D-branes with $D+d=9$ in a toroidal background, because it is expected that the spacetime-filling branes provide rich structure for the formation of lower-dimensional D-branes as we will mention below.}. We must consider gravity coupled to non-BPS D9-branes in order to analyze a time evolution of the universe.

There has recently been a great deal of interest in brane inflation \cite{inflation1} \cite{inflation2} \cite{inflation3} (for a review see, e.g., Ref. \cite{Gibbons}). If we consider Einstein gravity in the presence of spacetime-filling branes, the universe expands in an inflationary manner, because the tension energy of these branes can provide an effective cosmological constant. These work has been done by using Dirac-Born-Infeld type action \cite{BornInfeld} \cite{roll}\footnote{Calcagni has investigated the brane inflation by using cubic string field theory in Ref. \cite{CSFTinflation}.}. However, since we have computed the finite temperature effective potential on the basis of BSFT, it is natural to deal with the tachyon field in the framework of BSFT. The time dependent background at zero temperature has been discussed by Sugimoto and Terashima in the framework of BSFT \cite{BSFTcosmo}. They have also argued the production of tachyon matter \cite{roll} after the decay of the non-BPS D$p$-brane in the framework of BSFT. It would be interesting to generalize their calculation to the finite temperature case. It is expected that, even if these branes are stable near the Hagedorn temperature initially, they become unstable because the energy density decreases as the universe expands. This implies that the initial conditions of a rolling tachyon are realized in the early universe. Namely, the tachyon field is placed on the top of the potential. Then the tachyon starts to roll down from the local maximum of the potential at $T=0$ \cite{roll}.

The spacetime-filling branes are very advantageous in the sense that all the lower-dimensional D-branes in type II string theory are realized as topological defects through tachyon condensation from non-BPS D9-branes and {\D{9}} pairs. We can identify the topological charge as the Ramond-Ramond charge of the resulting D-branes \cite{nonBPSD} \cite{Senconjecture}. These D-brane charges can be classified using K-theory \cite{Ktheory1} \cite{Ktheory2}. Thus, if non-BPS D9-branes exist in the early universe, various kinds of branes may form through tachyon condensation \cite{CosCre}. It would be interesting to examine the possibility that our Brane World forms as a topological defect in a cosmological context \cite{inflation3} \cite{CosCre} \cite{CreUni} \cite{Warped} \cite{WhyLive}. In this paper, we study the homogeneous and isotropic tachyon condensation as a first step towards `Brane World Formation Scenario'\footnote{In this paper, we investigate the simplest case in the presence of the non-BPS D9-branes, namely, the homogeneous and isotropic (9+1)-dimensional spacetime case, and leave the construction of our $(3+1)$-dimensional Brane World for future work.}.

This paper is organized as follows. In \S \ref{sec:Action} we describe the action we consider. We investigate the time evolution of the universe in the high temperature case and in the zero temperature case in order to construct a cosmological scenario in the following two sections. We calculate the cosmological solution by using Einstein gravity in \S \ref{sec:dilfix}, and by using dilaton gravity in \S \ref{sec:Eindilgra}. We conclude in \S \ref{sec:conclusion} with a discussion of future directions. We have also included an appendix, in which we shortly mention the Minahan-Zwiebach model case.

\section{Action}
\label{sec:Action}

We will begin with describing the action for gravitational field. The low energy effective action for tree level closed strings is described by type IIA supergravity. For simplicity, we shall focus on 10-dimensional metric $g_{\mu \nu}$ and dilaton $\phi$, and set the other fields to zero. Then the action is given by
\BE
  S_{dil} = - \ \frac{1}{2 \kappa^2} \int d^{10} x \sqrt{-g} \ e^{-2 \phi}
    \left( {\cal R} + 4 \nabla_{\mu} \phi \nabla^{\mu} \phi \right),
\label{eq:dilaction}
\EE
where the constant $\kappa$ is defined as
\BE
  \kappa^2 = \frac{1}{2} \ (2 \pi)^7 {\ap}^4,
\EE
and ${\cal R}$ denotes the scalar curvature. We will investigate the time evolution of the universe in the string frame by using this action in \S \ref{sec:Eindilgra}. If we consider the constant dilaton case, then the action is given by usual Einstein-Hilbert one
\BE
  S_E = - \ \frac{1}{2 \kappa^2} \int d^{10} x \sqrt{-g} \ {\cal R}.
\label{eq:Einsteinaction}
\EE
We can apply this action when the dilaton has settled down to the (local) minimum of the dilaton potential, which is expected to be generated by a non-perturbative effect in string theory \cite{dilpot}. We will investigate the time evolution of the universe by using this action in \S \ref{sec:dilfix}.

We must also consider the action for non-BPS D9-branes. For simplicity, we only deal with the high temperature case and the zero temperature case. Although we have derived the free energy formally at any temperature, it is very complicated function at the intermediate temperature \cite{Hotta4} \cite{Hotta5} \cite{Hotta6}. First, let us describe the high temperature case. If there are $N$ non-BPS D9-branes, the tachyon field $\TM$ is an $N \times N$ real matrix in the adjoint representation of the $U(N)$ gauge group \cite{nonBPSD} as we have mentioned in the previous section. In the case that the universe is sufficiently hot and the non-BPS D9-branes are stable, $\TM = 0$ is the potential minimum \cite{Hotta6}. It is sufficient to deal with the action for open string gas because the total energy is dominated by open string gas if we consider the thermodynamic balance between open string gas and closed string one \cite{Hotta6}. In this case the action is represented as
\BE
  S_{gas} = \int d^{10} x \sqrt{-g} \ F(\B \sqrt{g_{00}}),
\label{eq:openstringgasaction}
\EE
where $F$ is the free energy of open string gas, and $\B$ is the inverse of the temperature. We have omitted the argument $T$ of the free energy $F$ because we are considering the case that the tachyon stays at the potential minimum $\TM = 0$. We will argue the cosmological solution by using this action in \S \ref{sec:dilfixstringgas} and \S \ref{sec:stringgas}. The energy-momentum tensor, which we can derive from this action, can be represented in a perfect fluid form such as
\BE
  {T_{\mu}}^{\nu} = diag (- \rho, p, \cdot \cdot \cdot , p).
\EE
by using energy density $\rho$ and pressure $p$. The entropy of open strings on non-BPS D9-branes near the Hagedorn temperature have been computed in \cite{Hotta6}. The result is
\BE
  S \simeq \BH E + \frac{2 N}{{\BH}^4} \sqrt{\frac{E \v{9}}{\pi}}.
\label{eq:opengasentropy}
\EE
From this we can compute the equation of state as
\BE
  p = \frac{1}{\B} \ \frac{\p S}{\p \v{9}} \simeq w \sqrt{\rho},
\label{eq:eqofstate1}
\EE
where the proportional constant $w$ is given by
\BE
  w = \frac{N}{\sqrt{\pi} {\BH}^5}.
\label{eq:eqofstateconst}
\EE
It is noteworthy that this equation of state is different from the usual one, which is represented as $p \propto \rho$. The approximation (\ref{eq:opengasentropy}) is valid only when the first term in the right hand side is much larger than the second term. This means that we can apply the equation of state (\ref{eq:eqofstate1}) if the energy density $\rho$ satisfies
\BE
  \rho \gg w^2.
\label{eq:Hagedorncondition}
\EE
Namely, we can apply the approximation when the energy density is much larger than the string scale. This condition is crucial when we argue the cosmological solution near the Hagedorn temperature.

Secondly, we describe the zero temperature case. Since we have computed the finite temperature effective potential for non-BPS D-branes on the basis of BSFT, it is consistent to argue the zero temperature case in the framework of BSFT. The BSFT action for a linear tachyon profile in the flat spacetime is derived in Ref. \cite{tachyon2}. Let us focus on tachyon $\TM$ in the open string spectrum, as well as graviton $g_{\mu \nu}$ and dilaton $\phi$ in the closed string one, and assume that the action in the curved spacetime is given by
\BE
  S_{dilT} = \mu_0 \int d^{10} x \sqrt{-g} \ 
    e^{- \phi} \ \textrm{Tr} \ e^{- \alpha {\TM}^2}
      \F \left( \lambda \nabla_{\mu} \TM \nabla^{\mu} \TM \right),
\label{eq:multitacdilaction}
\EE
where $\alpha$ and $\lambda$ are constants. If we follows the notation of Ref. \cite{Ohmori}, $\alpha = 1/4$ and $\lambda = \ap \ln 2$. $\mu_0$ is the constant part of the tension of a single non-BPS D9-brane, namely,
\BE
  \mu_0 = \frac{\sqrt{2}}{(2 \pi)^9 {\ap}^5},
\EE
and the function $\F (z)$ is defined as
\BE
  \F (z) = \frac{\sqrt{\pi} \ \Gamma (z+1)}
    {\Gamma \left( z + \frac{1}{2} \right)}.
\EE
Here we include the factor $e^{- \phi}$ as a contribution of dilaton since this action is derived from the tree level (disk) amplitude of open strings. Strictly speaking, we have not succeeded in constructing the action for open string tachyon field coupled to closed string modes. In addition, we have dropped the terms which include second and higher derivative of the tachyon field. There is room for argument on these points. We expect that the qualitative behavior of the time evolution of the universe can be described by this action. For simplicity, let us consider the case that $\TM$ is given by (\ref{eq:TM}). Then the action can be rewritten as
\BE
  S_{dilT} = \mu \int d^{10} x \sqrt{-g} \ 
    e^{- \phi} e^{- \alpha T^2}
      \F \left( \lambda \nabla_{\mu} T \nabla^{\mu} T \right),
\label{eq:tacdilaction}
\EE
where we have defined $\mu = \mu_0 N$. We will argue the cosmological solution by using this action in \S \ref{sec:dilrollingT}. If we assume that the dilaton is a constant, the action is given by
\BE
  S_T = \mu \int d^{10} x \sqrt{-g} \ e^{- \alpha T^2}
    \F \left( \lambda \nabla_{\mu} T \nabla^{\mu} T \right),
\label{eq:tacaction}
\EE
where the factor $e^{- \phi}$ can be absorbed into a redefinition of $\mu$. We will argue the cosmological solution by using this action in \S \ref{sec:dilfixrollingT}. Sugimoto and Terashima have shown that tachyon matter is produced by the decay of the non-BPS D$p$-brane in the flat spacetime and investigated the qualitative behavior of the cosmological solution by using this action \cite{BSFTcosmo}. The tachyon matter is pressureless gas with non-zero energy density, that is, its equation of state is represented as $p=0$ \cite{roll}. We will discuss the production of the tachyon matter in our cosmological model. If the derivative of $T$ is very small, then by using the approximation formula
\BE
  \F (z) \simeq 1+ (2 \ln 2) z,
\EE
under the condition $z \ll 1$, we obtain the action in the Minahan-Zwiebach model \cite{tachyon1}
\BE
  S_{MZ} = \mu \int d^{10} x \sqrt{-g} \ 
    e^{- \alpha T^2} ( \lambda \nabla_{\mu} T \nabla^{\mu} T + 1).
\label{eq:MZaction}
\EE
Sugimoto and Terashima have pointed out that $T$ diverges within finite time if we suppose this action \cite{BSFTcosmo}. We will shortly mention the numerical calculation in the Minahan-Zwiebach model case in appendix \ref{sec:MZmodel}.

\section{Constant Dilaton Case}
\label{sec:dilfix}

We investigate the time evolution of the universe in the constant dilaton case as a simple example of gravity coupled to the open string gas or the open string tachyon in this section. Although the constant dilaton is not a solution of dilaton gravity case as we will explain in the next section, we can apply this calculation if the dilaton has settled down to the (local) minimum of the dilaton potential as we have mentioned in the previous section.

\subsection{Open String Gas Case}
\label{sec:dilfixstringgas}

\begin{figure}
\begin{center}
$${\epsfxsize=6.5 truecm \epsfbox{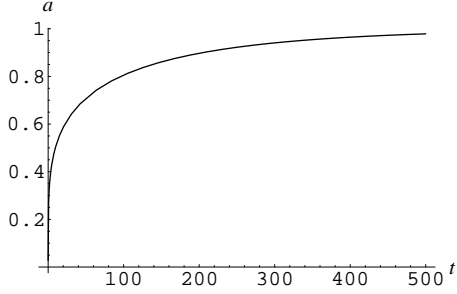}}$$
\caption{The time evolution of the scale factor $a(t)$ in the case of Einstein gravity coupled to open string gas. We have omitted the factor $( E_0 / w^2 )^{1/9}$ and have set $c=0$.}
\label{fig:aEingas}
\end{center}
\end{figure}

Let us first consider the high temperature case. In this case, it is sufficient to consider the action for open string gas for the system of strings and non-BPS D9-branes. Previously Einstein gravity coupled to open string gas on BPS D-branes has been investigated in the context of inflation due to the winding modes of open strings \cite{Haginflation}. We will show different type of solution. The action we consider here is the sum of the Einstein-Hilbert action (\ref{eq:Einsteinaction}) and that for the open string gas (\ref{eq:openstringgasaction}). In this case, the equation of motion is simply given by the Einstein equation
\BE
  {\cal R}_{\mu \nu} - \frac{1}{2} \ g_{\mu \nu} {\cal R}
    - \kappa^2 T_{\mu \nu} = 0.
\label{eq:Einsteineq}
\EE
Let us assume that the universe is spatially homogeneous and isotropic, for simplicity. We also suppose that the spatial curvature is flat. Then the line element is that of the spatially flat Robertson-Walker metric
\BE
  ds^{2} = - dt^{2} + a^2 (t) \sum_{i=1}^{9} (dx^{i})^2,
\label{eq:lineelement1}
\EE
where $a(t)$ is a scale factor and is a function of time $t$. We also suppose that the energy density $\rho$ and the pressure $p$ are functions of $t$,
\BA
  \rho &=& \rho (t),
\label{eq:homoisorho} \\
  p &=& p(t).
\label{eq:homoisop}
\EA
Under these conditions, the Einstein equation (\ref{eq:Einsteineq}) can be rewritten as
\BA
  36 H^2 - \kappa^2 \rho &=& 0, \\
  8 \dot{H} + 36 H^2 + \kappa^2 p &=& 0,
\EA
where $H$ is the Hubble parameter
\BE
  H= \frac{\dot{a}}{a},
\EE
and dot denotes the time derivative. From these equations, we can derive a modified energy conservation equation
\BE
  \dot{E} + 9HP = 0,
\label{eq:energyconservation}
\EE
where $E$ and $P$ are defined as
\BA
  E \equiv \rho a^9,
\label{eq:totalE} \\
  P \equiv p a^9.
\label{eq:volumeP}
\EA
$E$ is the total energy of open string gas. Substituting the equation of state for open string gas (\ref{eq:eqofstate1}) into (\ref{eq:energyconservation}), we obtain
\BE
  E = ({E_0}^{\frac{1}{2}} - w a^{\frac{9}{2}})^2,
\label{eq:Eexact}
\EE
where the constant $E_0$ equals to the total energy $E$ if the scale factor $a$ vanishes. We can easily derive the general solution for equations of motion by using these equations. The expansion solution is given by
\BE
  a (t) = \left( \frac{E_0}{w^2} \right)^{\frac{1}{9}}
    \left[ 1- \exp \left( - \ \frac{3w \kappa}{4} \ t + c \right)
      \right]^{\frac{2}{9}},
\label{eq:aexact}
\EE
where $c$ is an integration constant. We display the time evolution of the scale factor $a(t)$ in Figure \ref{fig:aEingas}. The universe evolves in the decelerated expansion phase at first. We cannot avoid the initial singularity in the framework of Einstein gravity coupled to open string gas. We must consider the $\ap$ correction near the curvature singularity. The scale factor asymptotically approaches to the constant value as $t \rightarrow \infty$. This is because the total energy $E$ asymptotically approaches to zero as we can see from (\ref{eq:Eexact}), and we are considering the case that the spatial curvature is zero. Thus, we cannot apply the equation of state (\ref{eq:eqofstate1}) in this region, because the condition (\ref{eq:Hagedorncondition}) is no longer satisfied in this case. The time reversal $t \rightarrow -t$ of (\ref{eq:aexact}) is also a solution and it represents a contraction solution.

\subsection{Rolling Tachyon Case}
\label{sec:dilfixrollingT}

Let us next consider the case that the tachyon rolls down the potential at zero temperature. The action we consider is the sum of Einstein-Hilbert one (\ref{eq:Einsteinaction}) and that for the tachyon field (\ref{eq:tacaction}). In this case, the equation of motion is given by
\BA
  {\cal R}_{\mu \nu} - \ \frac{1}{2} \ g_{\mu \nu} {\cal R}
    + \mu \kappa^2 e^{- \alpha T^2} g_{\mu \nu}
      \F (\lambda \nabla_{\alpha} T \nabla^{\alpha} T) \hspace{40mm} && \no \\
  - 2 \mu \kappa^2 e^{- \alpha T^2}
    \lambda \nabla_{\mu} T \nabla_{\nu} T
      \F' (\lambda \nabla_{\alpha} T \nabla^{\alpha} T) &=& 0, \\
  2 \lambda^2 \nabla_{\mu} \nabla_{\beta} T
    \nabla^{\beta} T \nabla^{\mu} T
      \F'' (\lambda \nabla_{\mu} T \nabla^{\mu} T) \hspace{50mm} && \no \\
  + \lambda \left( \nabla_{\mu} \nabla^{\mu} T
    - 2 \alpha T \nabla_{\mu} T \nabla^{\mu} T \right)
      \F' (\lambda \nabla_{\mu} T \nabla^{\mu} T) \hspace{10mm} && \no \\
  + \alpha T \F (\lambda \nabla_{\mu} T \nabla^{\mu} T) &=& 0.
\EA
We assume that the universe is homogeneous and isotropic like in the previous subsection, and that the tachyon $T$ is a function of $t$
\BA
  T = T(t),
\label{eq:tac}
\EA
as well as the line element is given by that of the spatially flat Robertson-Walker metric (\ref{eq:lineelement1}). Then the equations of motion are rewritten as
\begin{figure}
\begin{center}
$${\epsfxsize=13 truecm \epsfbox{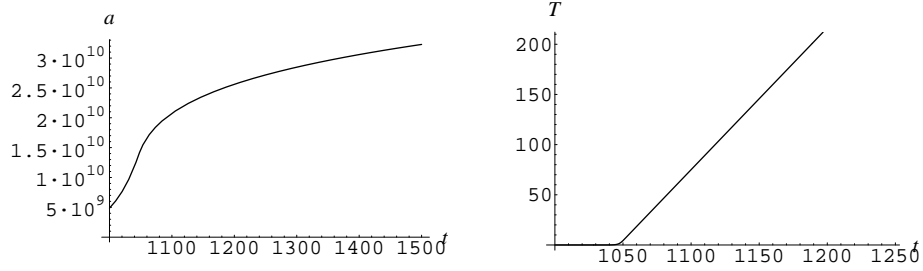}}$$
\caption{The time evolution of the scale factor $a(t)$ (left) and the tachyon field $T(t)$ (right) in the case of Einstein gravity coupled to tachyon field. We have set $a_0 = N = 1$. The initial condition for the tachyon is $T = 10^{-10}$ and $\dot{T} = 0$ at $t = 1000$. We choose such a small initial value of $T$ in order to show the inflation phase exists before the decelerated expansion phase.}
\label{fig:BSFTE}
\end{center}
\end{figure}
\BA
  36 H^2 - \mu \kappa^2 e^{- \alpha T^2} \F (- \lambda \dot{T}^2)
    - 2 \mu \kappa^2 e^{- \alpha T^2} \lambda \dot{T}^2
      \F' (- \lambda \dot{T}^2) &=& 0,
\label{eq:BSFTeqofmo1} \\
  8 \dot{H} + 36 H^2
    - \mu \kappa^2 e^{- \alpha T^2} \F (- \lambda \dot{T}^2) &=& 0,
\label{eq:BSFTeqofmo2} \\
  2 \lambda^2 {\dot{T}}^2 \ddot{T} \F''(- \lambda {\dot{T}}^2)
    - \lambda \left( \ddot{T} +9 H \dot{T}
      - 2 \alpha T {\dot{T}}^2 \right) \F'(- \lambda {\dot{T}}^2)
        \hspace{10mm} && \no \\
  + \alpha T \F(- \lambda {\dot{T}}^2) &=& 0.
\label{eq:BSFTeqofmo3}
\EA
There are three equations although we have only two unknown functions $a(t)$ (or $H(t)$) and $T(t)$. However, we can show that independent equations are only two equations.

In order to perform the numerical calculation, we must choose a reasonable initial condition. We have supposed that the non-BPS D9-branes are stable and $T=0$ in the high temperature case. We can easily solve the equations of motion at zero temperature if the tachyon remains at $T=0$. Substituting $T=0$ into above equations of motion, we obtain the de Sitter solution
\BA
  a = a_0 \exp \left( \frac{1}{6} \ \mu^{\frac{1}{2}} \kappa t \right),
\EA
where the constant $a_0$ is the scale factor at $t=0$. This is because the tension energy of these branes can provide an effective cosmological constant. Thus, it is reasonable to choose the initial condition which is close to the de Sitter solution. We calculate the numerical solution by choosing (\ref{eq:BSFTeqofmo2}) and (\ref{eq:BSFTeqofmo3}) as the independent equations as is depicted in Figure \ref{fig:BSFTE}. From this we can see that the tachyon asymptotically approaches to a linear function of $t$. Sugimoto and Terashima have pointed out that the tachyon asymptotically approaches to
\BE
  T \rightarrow \frac{t}{\sqrt{\lambda}} + \textrm{const}.
\label{eq:BSFTTlinear}
\EE
as $t \rightarrow \infty$ \cite{BSFTcosmo}. This comes from the divergence of $\F (z)$ and its derivative at $z = -1$. We can derive (\ref{eq:BSFTTlinear}) from $- \lambda {\dot{T}}^2 = -1$. Sugimoto and Terashima have shown that the ratio $p / \rho$ for the tachyon field vanishes at $z = -1$. Thus, tachyon matter is produced in this case. As we can see from Figure \ref{fig:BSFTE}, the scale factor asymptotically approaches to a constant as $t \rightarrow \infty$. This is because the energy density of the tachyon field approaches to zero as $t \rightarrow \infty$ and we are considering the case that the spatial curvature is zero. The inflation phase continues for a long time if we choose large $N$ or small initial value of $T$.

The whole story in the constant dilaton case is summarized as follows. The universe evolves in the decelerated expansion phase near the Hagedorn temperature at first. In this phase, $T=0$ is the potential minimum and Non-BPS D9-branes are stable. The temperature decreases as the universe expands. Then $T=0$ becomes the local maximum of the potential and the universe expands inflationary. The tachyon starts to roll down the potential and the non-BPS D9-branes decay. Finally, the universe turns to be in the decelerated expansion phase.

\section{Dilaton Gravity Case}
\label{sec:Eindilgra}

We have discussed the constant dilaton case in the previous section. However, type IIA supergravity contains dilaton gravity and the constant dilaton is not a solution in the dilaton gravity without dilaton potential as we will see below. Even if the dilaton potential exists, there is no reason why the dilaton settles down to the potential minimum from the beginning of the universe. In this sense it is natural to consider the scenario that the dilaton starts from the any point in the dilaton potential, then it moves towards the potential minimum, and finally it settles down to the potential minimum. As a first step towards such a scenario, we will consider the dilaton gravity case without the dilaton potential in this section.

\subsection{Open String Gas Case}
\label{sec:stringgas}

Let us first consider the high temperature case. The action we consider is the sum of that for dilaton gravity (\ref{eq:dilaction}) and that for open string gas (\ref{eq:openstringgasaction}). In this case, the equation of motion in the string frame is calculated as
\BA
  2 {\cal R}_{\mu \nu} +4 \nabla_{\mu} \nabla_{\nu} \phi
    -2 \kappa^2 e^{2 \phi} T_{\mu \nu} &=& 0,
\label{eq:eqofmotionEindilgas1} \\
  {\cal R} - 4 \nabla_{\mu} \phi \nabla^{\mu} \phi
    +4 \nabla_{\mu} \nabla^{\mu} \phi &=& 0.
\label{eq:eqofmotionEindilgas2}
\EA
We assume that the universe is homogeneous and isotropic like in the previous section, and that the dilaton $\phi$ is a function of $t$,
\BE
  \phi = \phi (t),
\label{eq:dilaton1}
\EE
as well as the line element, the energy density $\rho$ and the pressure $p$ are given by (\ref{eq:lineelement1}) $\sim$ (\ref{eq:homoisop}), respectively. Then the equations of motion are rewritten as
\BA
  9 (\dot{H} +H^2) -2 \ddot{\phi} + \kappa^2 e^{2 \phi} \rho &=& 0, \\
  \dot{H} +9H^2 -2H \dot{\phi} - \kappa^2 e^{2 \phi} p &=& 0, \\
  9 (\dot{H} +5H^2) +2 {\dot{\phi}}^2 -2 \ddot{\phi} -18H \dot{\phi} &=& 0.
\EA
Since these equations of motion are invariant under the T-duality transformation \cite{Tdual} \cite{scaledual}, it is convenient to introduce the new variable $ \varphi$ as
\BE
  \varphi \equiv 2 \phi -9 \ln a.
\EE
Then the equations of motion can be rewritten as
\BA
  9 H^2 - \ddot{\varphi} + \kappa^2 E e^{\varphi} = 0,
\label{eq:eqsmotionEindilgashomo1} \\
  \dot{H} - H \dot{\varphi} - \kappa^2 P e^{\varphi} = 0,
\label{eq:eqsmotionEindilgashomo2} \\
  {\dot{\varphi}}^2 - 2 \ddot{\varphi} + 9 H^2 = 0,
\label{eq:eqsmotionEindilgashomo3}
\EA
where $E$ and $P$ are defined as (\ref{eq:totalE}) and (\ref{eq:volumeP}), respectively. We can derive a modified energy conservation equation (\ref{eq:energyconservation}) from these equations. Substituting the equation of state for open string gas (\ref{eq:eqofstate1}) into (\ref{eq:energyconservation}), we can also obtain the total energy $E$ as a function of the scale factor $a$ as (\ref{eq:Eexact}).

\begin{figure}
\begin{center}
$${\epsfxsize=13 truecm \epsfbox{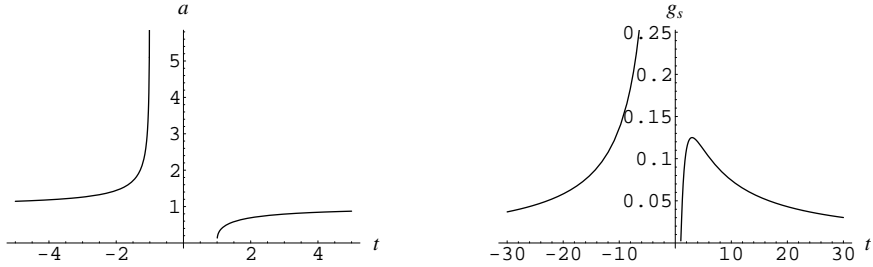}}$$
\caption{The time evolution of the scale factor $a(t)$ (left) and the string coupling $g_s(t)$ (right) in the case of dilaton gravity coupled to open string gas. We have set $a_0 = b_1 = b_2 = d = 1$.}
\label{fig:P0}
\end{center}
\end{figure}

We must solve equations (\ref{eq:eqsmotionEindilgashomo1}) $\sim$ (\ref{eq:eqsmotionEindilgashomo3}) in order to obtain $a(t)$ and $\phi(t)$ (or $H(t)$ and $\varphi(t)$). However, it is difficult to compute the exact solution for the above non-linear differential equations. Recall that we can apply the equation of state (\ref{eq:eqofstate1}) when the energy density satisfies the condition (\ref{eq:Hagedorncondition}), that is, it is much larger than the string scale, as we have mentioned in \S \ref{sec:Action}. From (\ref{eq:Eexact}), this condition is satisfied only when
\BE
  {E_0}^{\frac{1}{2}} \gg w a^{\frac{9}{2}},
\EE
since $w$ is close to the string scale energy density, and $E$ is approximated as
\BE
  E \simeq E_0.
\EE
Thus, the total energy is almost a constant. Substituting this into (\ref{eq:energyconservation}), we find out that
\BE
  P \simeq 0.
\label{eq:eqofstate2}
\EE
Previously, Tseytlin and Vafa have derived the cosmological solution in the case that $P=0$ \cite{Tseycos}. They investigated such a case because the equation of state of closed string gas can be approximated as (\ref{eq:eqofstate2}). In this case we can obtain the general solution. One of the solutions is given by
\BA
  a(t) &=& a_0 \left( \frac{t - b_2}{t + b_1} \right)^{\frac{1}{3}}, \\
  \phi (t) &=& \ln \ \frac{{a_0}^9 \mid t - b_2 \mid}{d (t + b_1)^2},
\label{eq:pressurezerodilaton}
\EA
where $a_0$, $d$, $b_1$ and $b_2$ are constants. $b_1$ and $b_2$ satisfy $b_1 + b_2 > 0$. $d$ is given by
\BE
  d = \frac{\kappa^2 E_0}{2}.
\EE
$a_0$ corresponds to the value of $a(t)$ at $t \rightarrow \pm \infty$. This solution is defined in the region $t < - b_1$ or $t > b_2$, and is an expansion one. Another solution is its time reversal $t \rightarrow -t$ and is a contraction one. We shall concentrate on the expansion solution. Since the dilaton $\phi$ in (\ref{eq:pressurezerodilaton}) is a function of $t$, the coupling of strings $g_s$ is also a function of $t$. It is given by
\BE
  g_s (t) = e^{\phi (t)}
    = \frac{{a_0}^9 \mid t - b_2 \mid}{d (t + b_1)^2}.
\EE
We display the time evolution of the scale factor $a(t)$ and the coupling $g_s (t)$ in Figure \ref{fig:P0}.

For $t < - b_1$, the universe evolves from the infinite past in the accelerated expansion phase like superinflation solution in the pre-big bang scenario \cite{GVcos}. It should be noted that there is no initial singularity. Then the scale factor and curvature diverge at $t = - b_1$. However, the temperature decreases as the scale factor increases, since the total energy $E$ is conserved. Then $T=0$ becomes the local maximum of the tachyon potential, and the tachyon starts to roll down the potential.

For $t \geq b_2$, on the other hand, the universe evolves in the decelerated expansion phase at first. We cannot avoid the initial singularity in this case. We must consider the $\ap$ correction near the curvature singularity. The scale factor asymptotically approaches to a constant like in the constant dilaton case in the previous section. If the integration constants satisfy ${E_0}^{\frac{1}{2}} \gg w {a_0}^{\frac{9}{2}}$, the universe remains constant volume, while if they satisfy ${E_0}^{\frac{1}{2}} \leq w {a_0}^{\frac{9}{2}}$, we cannot apply the equation of state (\ref{eq:eqofstate1}) or (\ref{eq:eqofstate2}) when the energy density no longer satisfies the condition (\ref{eq:Hagedorncondition}). The tachyon rolls down the potential in the latter case. We will investigate the rolling tachyon case in the next subsection.

\subsection{Rolling Tachyon Case}
\label{sec:dilrollingT}

Let us next consider the case that the tachyon rolls down the potential at zero temperature. The action we consider is the sum of that for dilaton gravity (\ref{eq:dilaction}) and that for the tachyon and the dilaton (\ref{eq:tacdilaction}). In this case, the equations of motion in the string frame is calculated as
\BA
  2 {\cal R}_{\mu \nu} + 4 \nabla_{\mu} \nabla_{\nu} \phi
    - \mu \kappa^2 e^{\phi - \alpha T^2}
      \left\{ 4 \lambda \nabla_{\mu} T \nabla_{\nu} T
        \F' (\lambda \nabla_{\alpha} T \nabla^{\alpha} T) \right.
          \hspace{10mm} && \no \\
  \left. - g_{\mu \nu} \F (\lambda \nabla_{\alpha} T \nabla^{\alpha} T)
    \right\} &=& 0, \\
  {\cal R} - 4 \nabla_{\mu} \phi \nabla^{\mu} \phi
    +4 \nabla_{\mu} \nabla^{\mu} \phi
      - \mu \kappa^2 e^{\phi - \alpha T^2}
        \F (\lambda \nabla_{\mu} T \nabla^{\mu} T) &=& 0, \\
  2 \lambda^2 \nabla_{\mu} \nabla_{\beta} T
    \nabla^{\beta} T \nabla^{\mu} T
      \F'' (\lambda \nabla_{\mu} T \nabla^{\mu} T) \hspace{60mm} && \no \\
  + \lambda \left( \nabla_{\mu} \nabla^{\mu} T
    - 2 \alpha T \nabla_{\mu} T \nabla^{\mu} T
      - \nabla_{\mu} \phi \nabla^{\mu} T \right)
        \F' (\lambda \nabla_{\mu} T \nabla^{\mu} T) \hspace{10mm} && \no \\
  + \alpha T \F (\lambda \nabla_{\mu} T \nabla^{\mu} T) &=& 0.
\EA
We assume that the universe is homogeneous and isotropic like in the previous section, and that the line element, the tachyon $T$ and the dilaton $\phi$ are given by (\ref{eq:lineelement1}), (\ref{eq:tac}) and (\ref{eq:dilaton1}), respectively. Then the equations of motion are rewritten as
\BA
  18 (\dot{H} +H^2) -4 \ddot{\phi}
    + \mu \kappa^2 e^{\phi - \alpha T^2}
      \left\{ 4 \lambda {\dot{T}}^2 \F'(- \lambda {\dot{T}}^2)
        + \F(- \lambda {\dot{T}}^2) \right\} &=& 0,
\label{eq:BSFTdileqofmo1} \\
  2 (\dot{H} +9H^2) -4H \dot{\phi}
    + \mu \kappa^2 e^{\phi - \alpha T^2} \F(- \lambda {\dot{T}}^2) &=& 0,
\label{eq:BSFTdileqofmo2} \\
  18 (\dot{H} + 5 H^2) + 4 {\dot{\phi}}^2 - 4 \ddot{\phi} - 36 H \dot{\phi}
    - \mu \kappa^2 e^{\phi - \alpha T^2} \F(- \lambda {\dot{T}}^2) &=& 0,
\label{eq:BSFTdileqofmo3} \\
  2 \lambda^2 {\dot{T}}^2 \ddot{T} \F''(- \lambda {\dot{T}}^2)
    - \lambda \left( \ddot{T} +9 H \dot{T} - 2 \alpha T {\dot{T}}^2
      - \dot{\phi} \dot{T} \right) \F'(- \lambda {\dot{T}}^2)
        \hspace{10mm} && \no \\
  + \alpha T \F(- \lambda {\dot{T}}^2) &=& 0.
\label{eq:BSFTdileqofmo4}
\EA
There are four equations although we have only three unknown functions $a(t)$ (or $H(t)$) $\phi(t)$ and $T(t)$. However, we can show that independent equations are only three equations under the condition $\dot{\phi} \neq 0$. This means that the constant dilaton is not a solution. 

In order to obtain the reasonable initial condition for numerical calculation, we derive the exact solution in the $T=0$ case, namely, when the tachyon remains at metastable point. In this case the independent equations of motion is obtained from (\ref{eq:BSFTdileqofmo1}) $\sim$ (\ref{eq:BSFTdileqofmo4}) as
\BA
  -36H^2 -2 {\dot{\phi}}^2 +18H \dot{\phi}
    + \mu \kappa^2 e^{\phi} &=& 0, \\
  4 \dot{H} - \ddot{\phi} + H \dot{\phi} &=& 0.
\EA
Now we derived these equations in the string frame. If we consider the same problem in the Einstein frame, we obtain Einstein gravity coupled to a scalar field with an exponential potential. Previously the exact solution in such a case is calculated in \cite{scalarcosmo}. We can apply the similar method to our case. In order to solve these equations, it is convenient to define the variables $u, v, \tau$ as
\BA
  \ln a - \frac{1}{4} \ \phi &=& \frac{1}{9} (u+v),
\label{eq:aphiuv1} \\
  \phi &=& \frac{4}{3} (v-u),
\label{eq:phiuv1} \\
  \frac{d \tau}{dt} &=& \frac{3 \mu^{\frac{1}{2}} \kappa}{4} \ 
    \exp \left[ - \ \frac{2}{3} \ (u-v) \right].
\label{eq:tauuv}
\EA
Then, using these variables, the equations of motion can be rewritten as
\begin{figure}
\begin{center}
$${\epsfxsize=13 truecm \epsfbox{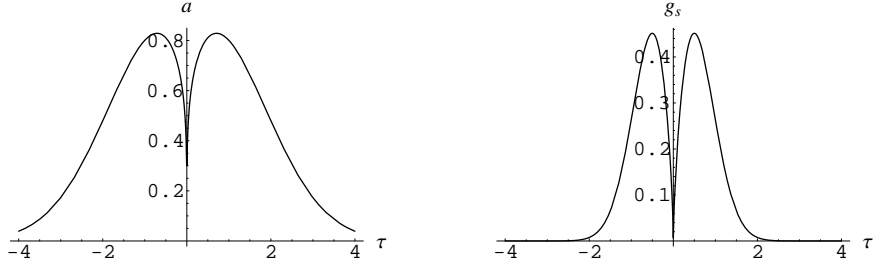}}$$
\caption{The time evolution of the scale factor $a(t)$ (left) and the string coupling $g_s(t)$ (right) in the $T=0$ case. We have set $c_1 = 1, \ \ \ c_2 = c_3 = c_4 = 0$.}
\label{fig:EindilT0}
\end{center}
\end{figure}
\BA
  & u' v' = 1,
\label{eq:uv1} \\
  & u'' + v'' + 2 (v')^2 - 2 u' v' = 0,
\label{eq:uv2}
\EA
where prime denotes the derivative by $\tau$. Eliminating $u$ from these equations, we obtain
\BA
  \left\{ 1 - \frac{1}{(v')^2} \right\} \left\{ v'' +2 (v')^2 \right\} = 0.
\EA
If we choose $(v')^2 = 1$, we obtain the constant dilaton solution, which does not satisfy the condition $\dot{\phi} \neq 0$. Thus, we have to choose
\BE
  v'' + 2 (v')^2 = 0.
\label{eq:v2v2}
\EE
Substituting (\ref{eq:uv1}) and (\ref{eq:v2v2}) into (\ref{eq:uv2}), we obtain
\BE
  u'' - 2 = 0.
\EE
We can easily derive the general solution of these equations as
\BA
  u &=& \tau^2 + c_3 \tau + c_4,
\label{eq:usol} \\
  v &=& \frac{1}{2} \ \ln | c_1 \tau + c_2 |,
\label{eq:vsol}
\EA
where $c_1 \sim c_4$ are integration constants. From (\ref{eq:uv1}), (\ref{eq:usol}) and (\ref{eq:vsol}), we can see that $c_1$, $c_2$ and $c_3$ are related as
\BA
  c_1 c_3 = 2 c_2.
\EA
Substituting (\ref{eq:usol}) and (\ref{eq:vsol}) into (\ref{eq:aphiuv1}) $\sim$ (\ref{eq:tauuv}), we obtain the solution in the $T=0$ case as
\BA
  \phi &=& \frac{2}{3} \left\{ \ln | c_1 \tau + c_2 |
    - 2 (\tau^2 + c_3 \tau + c_4) \right\}, \\
  a &=& | c_1 \tau + c_2 |^{\frac{2}{9}}
    \exp \left[ - \ \frac{2}{9} (\tau^2 + c_3 \tau + c_4) \right], \\
  \frac{d \tau}{dt} &=& \frac{3 \mu^{\frac{1}{2}} \kappa}{4}
    | c_1 \tau + c_2 |^{\frac{1}{3}}
      \exp \left[ - \ \frac{2}{3} (\tau^2 + c_3 \tau + c_4) \right],
\label{eq:dtaudt}
\EA
by using the parameter $\tau$. Since the right hand side of the last equation (\ref{eq:dtaudt}) is positive in $\tau < - c_2 / c_1$ and $\tau > - c_2 / c_1$, $t$ is a monotone increasing function of $\tau$ in both regions. We display the time evolution of the scale factor $a(t)$ and the coupling $g_s (t)$ in Figure \ref{fig:EindilT0}. For $\tau \leq - c_2 / c_1$, the universe evolves in the accelerated expansion phase at first and turns to be in the decelerated expansion phase, then contracts towards a singularity, while for $\tau \geq - c_2 / c_1$ the universe evolves in the decelerated expansion phase at first and turns to be in the contraction phase. The accelerated expansion solution in the high temperature case naturally connects to the accelerated expansion phase in $\tau \leq - c_2 / c_1$, while decelerated expansion solution in the high temperature case to the decelerated expansion phase in $\tau \geq - c_2 / c_1$.
\begin{figure}
\begin{center}
$${\epsfxsize=13 truecm \epsfbox{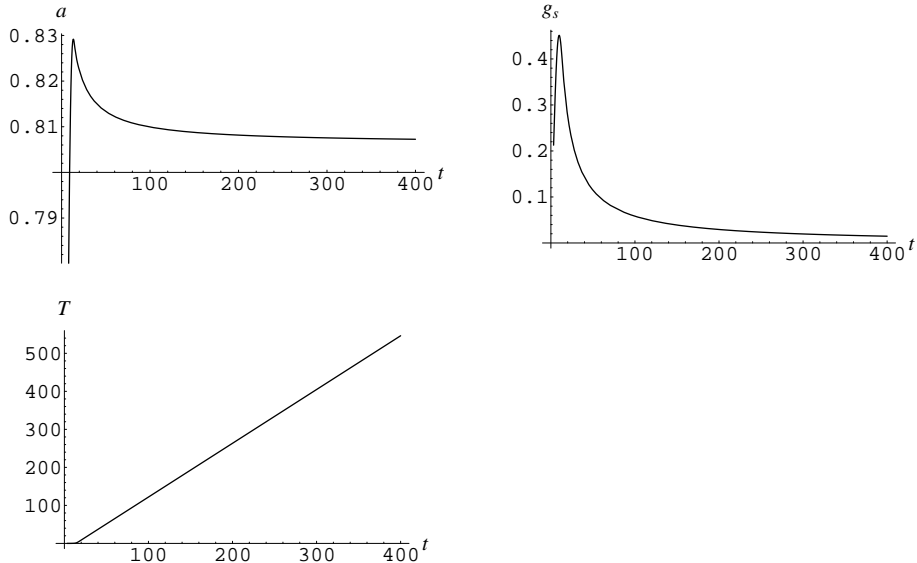}}$$
\caption{The time evolution of the scale factor $a(t)$ (left), the string coupling $g_s(t)$ (right) and the tachyon $T(t)$ (below) for $c_1 = 1$. We have set $N=1, \ \ \ c_2 = c_3 = c_4 = 0$. The initial condition for the tachyon is $T = 0.01$ and $\dot{T} = 0$ at $\tau = 0.1$.}
\label{fig:DecExp1}
\end{center}
\end{figure}

Before turning to the numerical calculation, we mention the time evolution of the universe after the tachyon rolls down the potential. In this case, tachyon matter is produced, since the tachyon asymptotically approaches to the linear function of $t$ (\ref{eq:BSFTTlinear}) as we will see below. The equation of motion for the tachyon matter is given by $p=0$. So, it is expected that the solution in the rolling tachyon case asymptotically approaches to the $P=0$ solution which we obtained in the high temperature case in the previous subsection. Therefore, if there is a solution which interpolate smoothly between the inflation phase in the $T=0$ case and the decelerated expansion solution in the $P = 0$ case, such a solution represent a completely non-singular cosmological model as has been proposed in the pre-big bang scenario \cite{GVcos}.

Let us now turn to the numerical calculation. From the equations of motion (\ref{eq:BSFTdileqofmo1}) $\sim$ (\ref{eq:BSFTdileqofmo4}), we can obtain
\begin{figure}
\begin{center}
$${\epsfxsize=13 truecm \epsfbox{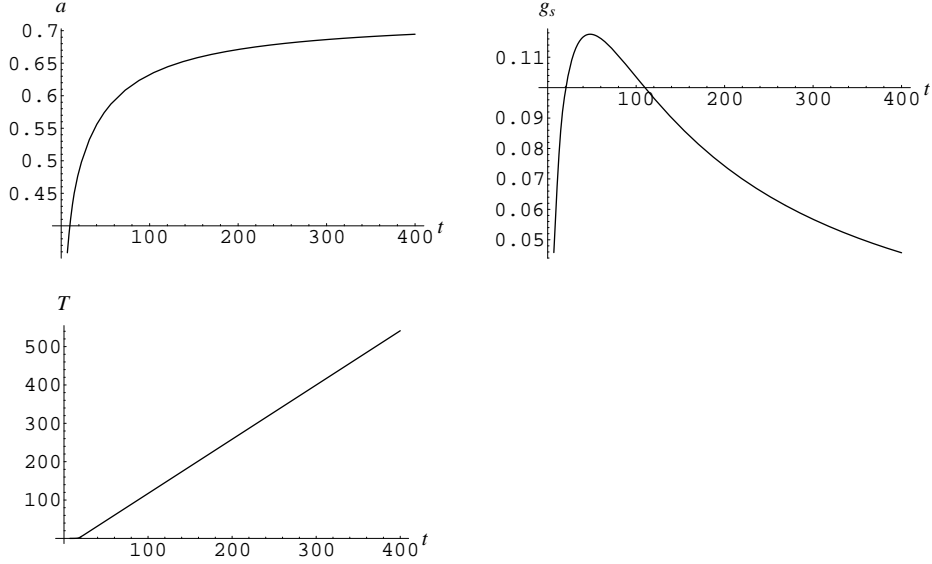}}$$
\caption{The time evolution of the scale factor $a(t)$ (left), the string coupling $g_s(t)$ (right) and the tachyon $T(t)$ (below) for $c_1 = 0.1$. We have set $N=1, \ \ \ c_2 = c_3 = c_4 = 0$. The initial condition for the tachyon is $T = 0.01$ and $\dot{T} = 0$ at $\tau = 0.1$.}
\label{fig:DecExp2}
\end{center}
\end{figure}
\BA
  \ddot{\phi} &=& {\dot{\phi}}^2 - 18 H^2
    - \ \frac{5}{2} \ \mu \kappa^2 e^{\phi - \alpha T^2}
      \F(- \lambda {\dot{T}}^2), \\
  \dot{H} &=& 2 H \dot{\phi} - 9 H^2 
    - \ \frac{1}{2} \ \mu \kappa^2 e^{\phi - \alpha T^2}
      \F(- \lambda {\dot{T}}^2), \\
  \ddot{T} &=& \left\{ 2 \lambda {\dot{T}}^2 \F''(- \lambda {\dot{T}}^2)
    - \F'(- \lambda {\dot{T}}^2) \right\}^{-1} \no \\
  && \times \left\{ \left( 9 H \dot{T} - 2 \alpha T {\dot{T}}^2
      - \dot{\phi} \dot{T} \right) \F'(- \lambda {\dot{T}}^2)
        - \ \frac{\alpha}{\lambda} \ T \F(- \lambda {\dot{T}}^2) \right\}.
\EA
We choose these equations as the independent equations for the numerical calculation. Let us first consider the initial condition close to the decelerated expansion phase for $\tau \geq - c_2 / c_1$ in the $T=0$ case. It is expected that the decelerated expansion solution in the high temperature case is smoothly connected to this phase. If we take a large $c_1$, the transition to the contraction phase occurs before the tachyon rolls down the potential. We display the time evolution of the scale factor $a(t)$, the coupling $g_s (t)$ and the tachyon $T(t)$ in the $c_1 = 1$ case in Figure \ref{fig:DecExp1}. We must set $c_1$ to a small value in order to keep the decelerated expansion phase. We display the time evolution of the scale factor $a(t)$, the coupling $g_s (t)$ and the tachyon $T(t)$ in the $c_1 = 0.1$ case in Figure \ref{fig:DecExp2}. Thus, if we take a small $c_1$, we can connect the decelerated expansion phase in the $T=0$ case to the decelerated expansion solution in the $P=0$ case. We can see from Figure \ref{fig:DecExp1} and Figure \ref{fig:DecExp2} that the tachyon asymptotically approaches to the linear function of $t$ (\ref{eq:BSFTTlinear}) as $t \rightarrow \infty$ in both cases, like in the Einstein gravity case. Thus, the tachyon matter is produced, and the equation of state for the tachyon field becomes $p=0$. Taking the results in the high temperature case into the consideration, we may be able to construct the cosmological scenario in which the universe evolves in the decelerated expansion phase at high temperature, and this phase is kept even at low temperature.
\begin{figure}
\begin{center}
$${\epsfxsize=13 truecm \epsfbox{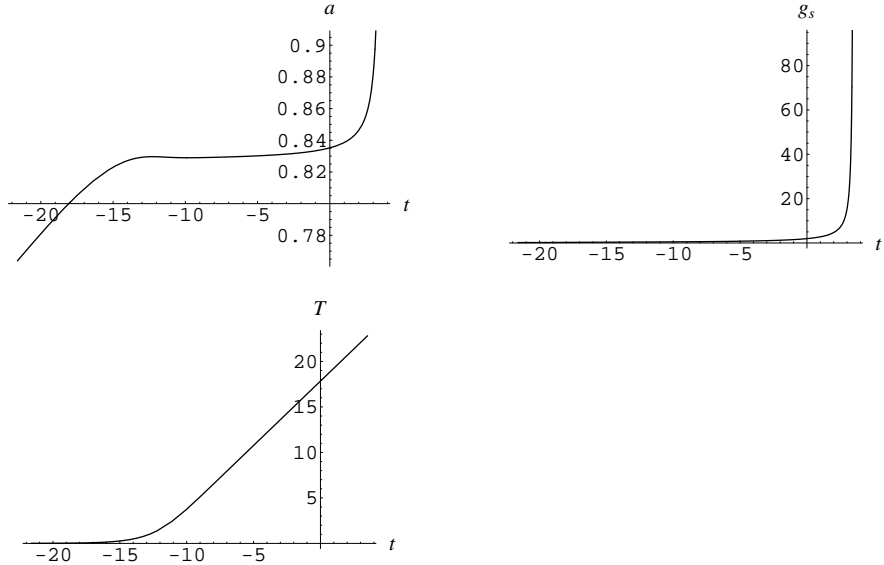}}$$
\caption{The time evolution of the scale factor $a(t)$ (left), the string coupling $g_s(t)$ (right) and the tachyon $T(t)$ (below) for the initial parameter $\tau = -1.17$. We have set $N = c_1 = 1, \ \ \ c_2 = c_3 = c_4 = 0$. The initial condition for the tachyon is $T = 0.01$ and $\dot{T} = 0$.}
\label{fig:Inflation1}
\end{center}
\end{figure}

\begin{figure}
\begin{center}
$${\epsfxsize=13 truecm \epsfbox{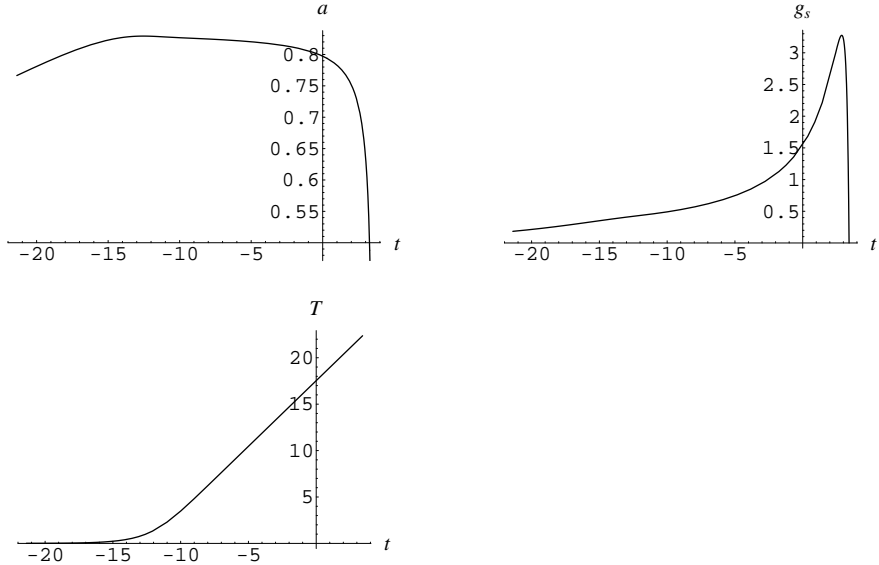}}$$
\caption{The time evolution of the scale factor $a(t)$ (left), the string coupling $g_s(t)$ (right) and the tachyon $T(t)$ (below) for the initial parameter $\tau = -1.16$. We have set $N = c_1 = 1, \ \ \ c_2 = c_3 = c_4 = 0$. The initial condition for the tachyon is $T = 0.01$ and $\dot{T} = 0$.}
\label{fig:Inflation2}
\end{center}
\end{figure}

Secondly, let us consider the initial condition close to the decelerated expansion phase for $\tau \leq - c_2 / c_1$ in the $T=0$ case. It is expected that the accelerated expansion solution in the high temperature case is smoothly connected to the accelerated expansion phase for $\tau \leq - c_2 / c_1$ in the $T=0$ case. The results are very sensitive to the initial condition. If we set the tachyon $T = 0.01$ and its derivative $\dot{T} = 0$ at $\tau = -1.17$ the universe evolves from deceleration to acceleration as is depicted in Figure \ref{fig:Inflation1}, while if we set them at $\tau = -1.16$ the universe evolves from expansion to contraction as is depicted in Figure \ref{fig:Inflation2}. The scale factor diverges or vanishes within finite time, and the curvature diverges in both cases. The same observation applies to the small $c_1$ case. Although it is not to be denied that there is a special initial condition which leads to a solution without curvature singularity, we obtain a solution with curvature singularity in general. It is difficult to avoid the curvature singularity within the framework of type IIA supergravity coupled to the tachyon field without the dilaton potential. We must deal with the $\alpha'$ corrections to the type IIA supergravity in these cases. We can see from Figure \ref{fig:Inflation1} and Figure \ref{fig:Inflation2} that the tachyon asymptotically approaches to the linear function of $t$ (\ref{eq:BSFTTlinear}) as $t \rightarrow \infty$ in both cases. Thus, the tachyon matter is produced also in these cases. Taking the results in the high temperature case into consideration, the universe evolves from the infinite past in the accelerated expansion phase, and the universe evolves in the accelerated expansion phase or in the accelerated contraction phase at last. We may be able to construct the cosmological model without the initial singularity, but the curvature diverges within finite time in the framework of dilaton gravity coupled to tachyon field, and we must consider the $\ap$ correction.

\section{Conclusion and Discussion}
\label{sec:conclusion}

We have investigated the time evolution of the universe in the presence of non-BPS D9-branes by using Einstein gravity and dilaton gravity. We try to construct the following scenario for the early universe: The universe expands near the Hagedorn temperature and the open string gas on the non-BPS D9-branes dominates the total energy of the system of strings and non-BPS D9-branes at first. The temperature decreases as the universe expands. Then $T=0$ becomes the local maximum of the potential and the non-BPS D9-branes become unstable at low temperature. Finally, the tachyon rolls down the potential and the non-BPS D9-branes disappear.

In the Einstein gravity case, the universe evolves in the decelerated expansion phase near the Hagedorn temperature at first, and turns to be in the inflation phase, and finally evolves in the decelerated expansion phase. We obtain two types of cosmological scenario in the dilaton gravity case. First, the universe evolves in the decelerated expansion phase at high temperature, and this phase is kept even at low temperature. Secondly, the universe evolves from the infinite past in the accelerated expansion phase, and the universe evolves in the accelerated expansion phase or in the accelerated contraction phase at last.

We have only dealt with the high temperature case and the zero temperature case. However, we must derive a solution at intermediate temperature in order to discuss our cosmological model more precisely\footnote{It must be noted that the tachyon field need not satisfy the slow-roll condition in the $(9+1)$-dimensional case. It is sufficient that the smoothness and flatness problems are resolved on our Brane World during its formation.}. In addition to this, we must consider $\ap$ correction for supergravity, higher derivative of the tachyon field, higher-loop correction, and so on. Much still remains to be done.

We have constructed our model by using dilaton gravity without the dilaton potential in this paper. However there is a possibility that, if we consider dilaton gravity with the dilaton potential, we can find out a solution which interpolate smoothly between the accelerated expansion phase and decelerated expansion one. It is worth while examining dilaton gravity with the dilaton potential and deriving the energy condition for graceful exit like it has been done in the case of the pre-big bang scenario \cite{GracefulExit}.

We have only considered the very simple case that the tachyon field is represented as (\ref{eq:TM}) and that the tachyon rolls down homogeneously. In general, we must deal with an arbitrary matrix which depends not only on the time coordinate $t$ but also on the spatial coordinate $x_i$. If the tachyon condensates in a topologically non-trivial configuration, lower dimensional branes form as topological defects as we have mentioned in \S \ref{sec:Intro}. We are able to argue the probability of formation of any type of branes if we can estimate the time evolution of the tachyon field in the gravitational background. In particular, it is worth while investigating the D3-brane case, since its world volume is 4-dimensional spacetime. It would be interesting to examine the possibility of cosmological brane models such as Randall-Sundrum model \cite{RS1} \cite{RS2}, Brane Gas Cosmology \cite{BGC}, ekpyrotic universe \cite{ekpyrotic}, KKLT model \cite{KKLT} and KKLMMT model \cite{KKLMMT} in this context \cite{inflation3} \cite{CosCre} \cite{CreUni} \cite{Warped} \cite{WhyLive}. 

We have discussed the production of tachyon matter, but ignored the creation of closed strings from the decaying non-BPS D9-branes. The creation of closed strings in the rolling tachyon background has been discussed recently \cite{closedcre}. Sen, Yi, Yee and Gutperle have emphasized that tachyon matter is nothing but the closed strings created at the end of brane decay based on the analysis of decay of non-BPS D-brane in the presence of electric field \cite{opencloseddual}. It would be interesting to investigate the effect of the closed strings to our cosmological model.

\section*{Acknowledgements}

The author would like to thank S. Sugimoto, S. Terashima, K. Hashimoto, H. Shimada, Y. Sato and colleagues at KEK and University of Tokyo, Komaba for useful discussions. He appreciates the Yukawa Institute for Theoretical Physics at Kyoto University. Discussions during the YITP workshop YITP-W-05-08 on "String Theory and Quantum Field Theory," YITP-W-05-09 on "Thermal Quantum Field Theories and Their Applications" and YITP-W-05-21 on "Fundamental Problems and Applications of Quantum Field Theory" were useful to complete this work.

\appendix

\section{Minahan-Zwiebach Model Case}
\label{sec:MZmodel}

\begin{figure}
\begin{center}
$${\epsfxsize=6.5 truecm \epsfbox{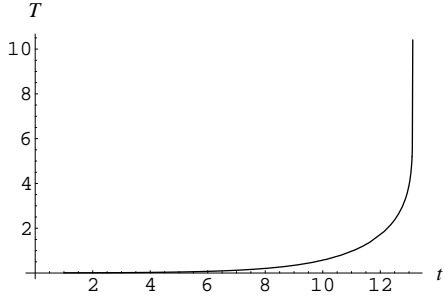}}$$
\caption{The time evolution of the tachyon $T$ in the case of Einstein gravity coupled to tachyon field in the Minahan-Zwiebach model. We have set $a_0 = N = 1$. The initial condition for the tachyon is $T = 0.01$ and $\dot{T} = 0$ at $t = 1$.}
\label{fig:MZmodel}
\end{center}
\end{figure}

Here, we shortly mention the Minahan-Zwiebach model case. The action we consider is the sum of that for Einstein-Hilbert one (\ref{eq:Einsteinaction}) and that for Minahan-Zwiebach model (\ref{eq:MZaction}). Then the equations of motion are given by
\BA
  {\cal R}_{\mu \nu} - \ \frac{1}{2} \ g_{\mu \nu} {\cal R}
    + \mu \kappa^2 e^{- \alpha T^2} g_{\mu \nu}
      \left( \lambda \nabla_{\alpha} T \nabla^{\alpha} T + 1 \right)
        - 2 \mu \kappa^2 e^{- \alpha T^2}
          \lambda \nabla_{\mu} T \nabla_{\nu} T &=& 0, \\
  \lambda \nabla_{\mu} \nabla^{\mu} T
    - \alpha \lambda T \nabla_{\mu} T \nabla^{\mu} T
      + \alpha T &=& 0.
\EA
We assume that the universe is homogeneous and isotropic, and that the line element and tachyon $T$ are given by (\ref{eq:lineelement1}) and (\ref{eq:tac}), respectively, as we have done so far. Then the equations of motion are rewritten as
\BA
  36 H^2 - \mu \kappa^2 e^{- \alpha T^2} ( \lambda \dot{T}^2 + 1 ) &=& 0, \\
  8 \dot{H} + 36 H^2
    + \mu \kappa^2 e^{- \alpha T^2} ( \lambda \dot{T}^2 - 1 ) &=& 0, \\
  \lambda \ddot{T} + 9 \lambda H \dot{T}
    - \alpha \lambda T \dot{T}^2 - \alpha T &=& 0.
\EA
We choose the initial condition close to de Sitter solution like in the \S \ref{sec:dilfixrollingT}. The result of numerical calculation is depicted in Figure \ref{fig:MZmodel}. The tachyon reaches to the potential minimum within finite time as Sugimoto and Terashima have pointed out \cite{BSFTcosmo}. We can also show that the tachyon diverges within finite time in the dilaton gravity case.

\end{document}